\newcommand{\teff}{T$_{\mathrm{eff}}$}

\newcommand{\logg}{log~$g$}
\newcommand{\feh}{[Fe/H]}
\newcommand{\kms}{km~s$^{-1}$}
\newcommand{\vsini}{V$\sin{i}$}

\newcommand{\Mgrad}{$\Delta$[Fe/H]/$\Delta$R}
\newcommand{\Vgrad}{$\Delta$[Fe/H]/$\Delta$Z}

\newcommand{\Msun}{M\textsubscript{\(\odot\)}}
\def\code#1{\texttt{#1}}

\newcommand{\numstar}{129 148} 

\documentclass[a4paper,fuseAMS,leqn,usenatbib,letter]{mnras}

\usepackage{graphicx}
\usepackage{amssymb}
\usepackage{amsmath}
\usepackage{times}
\usepackage{subfigure}
\usepackage{float}
\usepackage{longtable}
\usepackage[T1]{fontenc}
\usepackage{ae,aecompl}

\title[Metallicity gradients in LAMOST and Gaia]{Chemical Cartography with LAMOST and Gaia Reveal Azimuthal and Spiral Structure in the Galactic Disk}

 \author[Hawkins 2022]{Keith~Hawkins$^{1}$\thanks{E-mail: keithhawkins@utexas.edu}   \\
$^{1}$Department of Astronomy, The University of Texas at Austin, 2515 Speedway Boulevard, Austin, TX 78712, USA\\
 }

\date{Accepted 2023 April 24. Received 2023 April 22; in original form 2022 July 11 }

\pubyear{2022}

\begin{document}
\label{firstpage}
\pagerange{\pageref{firstpage}--\pageref{lastpage}}
\maketitle

\begin{abstract}
Chemical Cartography, or mapping, of our Galaxy has the potential to fully transform our view of its structure and formation. In this work, we use chemical cartography to explore the metallicity distribution of OBAF-type disk stars from the LAMOST survey and a complementary sample of disk giant stars from Gaia DR3. We use these samples to constrain the radial and vertical metallicity gradients across the Galactic disk. We also explore whether there are detectable azimuthal variations in the metallicity distribution on top of the radial gradient.  For the OBAF-type star sample from LAMOST, we find a radial metallicity gradient of  \Mgrad\ $\sim -0.078 \pm 0.001$~dex/kpc in the plane of the disk and a vertical metallicity gradient of   \Vgrad\ $\sim -0.15 \pm 0.01$~dex/kpc in the solar neighborhood. The radial gradient becomes shallower with increasing vertical height while the vertical gradient becomes shallower with increasing Galactocentric radius, consistent with other studies. We also find detectable spatially-dependent azimuthal variations on top of the radial metallicity gradient at the level of $\sim$0.10~dex. Interestingly, the azimuthal variations appear be close to the Galactic spiral arms in one dataset (Gaia DR3) but not the other (LAMOST). These results suggest that there is azimuthal structure in the Galactic metallicity distribution and that in some cases it is co-located with spiral arms. 
\end{abstract}


\section{Introduction}
\label{sec:Introduction}
Understanding the formation, assembly, and evolution of galaxies across the universe is a key goal of astrophysics. To this end, the chemical, dynamical, and spatial properties of the constituent stars of galaxies are key tracers of the physics process that govern galaxy formation. Our own Galaxy represents an excellent laboratory for understanding galaxy formation and assembly because it is one of the few systems where we can (and have) obtain(ed) high-precision positions, velocities, kinematics, and chemical abundances for millions of stars. With these data, there has been a large breadth of literature on the chemical and dynamical nature for a large number of stars in our Galaxy, which has been used to infer how our Galaxy came to be. There has also been an explosion in our understanding of the distribution of stars and their properties across the Galaxy that has been facilitated by an industrial revolution in stellar spectroscopy. These data have enabled chemical cartography, or mapping, of the Milky Way through its stars \citep[e.g.,][]{Hayden2015}.  One of result from chemical cartographic studies, which confirmed earlier work \citep[e.g.,][]{Mayor1976},  is that both global metallicity, and other chemical abundances, have spatial variations across the Milky Way's disk in the form of gradients \citep[e.g.][]{Boeche2013, Cunha2016}. These gradients have enabled us to not only learn about the structure of the Galaxy but have also provided key insights to its `inside-out' formation history. 

It has been shown \citep[e.g.,][and references therein]{Andrievsky2002, Hou2002, Friel2002, Nordstrom2004, AllendePrieto2006, Lemasle2008, Magrini2009, Wu2009, Pedicelli2009, Friel2010, Ruchti2011, Luck2011, Bilir2012, Boeche2013b, Boeche2014, RecioBlanco2014, Mikolaitis2014, Pleven2015, Xiang2015,  Jacobson2016, Cunha2016, Onal2016, Netopil2016,Yan2019, GaiaCC22} that the Milky Way's disk (whether thick or thin), has a negative radial metallicity gradient\footnote{In this work, we define the radial metallicity gradient as the change in the metallicity of a population of stars divided by the change in their Galactocentric radius (i.e., \Mgrad). Alternatively, the vertical metallicity gradient, \Vgrad, is defined as the change in the metallicity of population of stars divided by the change their absolute vertical height (|Z|). } such that the inner Galaxy (where the Galacticocentirc radius, R, is less than the solar value)  is, on average, more metal-rich compared to the outer disk (where  R $\geq$ 8.27~kpc). This negative radial metallicity gradient ranges from  --0.10 $<$ \Mgrad\ $<$ 0.00 ~dex/kpc in the Galactic plane \citep[e.g.][and references therein]{Onal2016} depending on the tracer population and survey volume. The negative radial metallicity gradient in the Galactic disk was strong evidence that the Galactic disk must have formed in an `inside-out' manner \citep[e.g.,][and references therein]{Frankel2019}, whereby the inner Galaxy formed early and fast and at later times the outer Galaxy formed. In addition to the radial metallicity gradient, many studies \citep[e.g.][and references therein]{Bartaviute2003, Karaali2003, AllendePrieto2006, Soubiran2008, Yaz2010, Ruchti2011, Kordopatis2011, Chen2011, Bilir2012, Hayden2014,Bergemann2014, Boeche2014, Huang2015, Pleven2015, Xiang2015, Yan2019, Nandakumar2020,GaiaCC22} have also found that the Galactic disk has a negative absolute vertical metallicity gradient. This gradient ranges between --0.25~$<$~\Vgrad~$<$~--0.10~dex/kpc. Thus as the tracer population of stars gets further above (or below) the Galactic mid-plane, the average metallicity decreases. The size of the vertical gradient varies significantly with Galactocentric radius, becoming smaller in the outer Galaxy compared to the inner Galaxy \citep[e.g.,][and references therein]{Onal2016}. 

The fact that there are radial and vertical metallicity gradients observed in the Galactic disk is established at this point, though there are some slight disagreements about the size of those gradients across the literature, likely related to the different samples/tracers used and volumes probed. That said, one area that requires more attention is to quantify whether there are azimuthal variations on top of the radial metallicity gradient. These azimuthal variations could arise by secular processes \citep[e.g.][]{Dimatteo2013, Khoperskov2018,Fragkoudi2018, Wheeler2021, Bellardini2022,Filion2023} or stellar migration due to tidal interactions between the Galactic disk and a Sagittarius-like dwarf galaxy \citep[e.g.][]{Carr2022}.  It may even be expected that azimuthal variations in metallicity could track along the spiral arms \citep[e.g.,][]{Khoperskov2021, Poggio2022}. Additionally, azimuthal variation in the radial metallicity gradient (and thus distribution) have also been observed in other Galaxies \citep[e.g.,][]{Hwang2019}. However, the presence (or lack thereof) and size of azimuthal variations in the metallicity gradients observed in our own Galaxy has been not well explored.  

Therefore, in this paper, we present the radial (\Mgrad) and vertical (\Vgrad) metallicity gradient, as measured by O-, B-, A- and early F-type stellar tracers observed in the LAMOST survey (sample I) along with a complementary sample from of giant stars observed within the third data release from the Gaia spacecraft \cite[Gaia DR3,][]{Gaiasummary2022, RecioBlanco2022} (sample II). In addition, we will also attempt to observationally constrain the level of any azimuthal variations that may lie on top of the metallicity gradient and discuss their implications. To facilitate the presentation of our results, this paper is outlined in the following way: in section~\ref{sec:data} we outline the where the data for this project is sourced from, i.e. LAMOST (section~\ref{subsec:selection}) and Gaia (sections~\ref{subsec:DR3RVS}, \ref{subsec:EDR3}) and the criteria used to ensure a high-quality sample. In section~\ref{sec:Methods}, we discuss the methods used to derive the metallicity gradients from the observable data. We present, in section~\ref{sec:result}, the primary results on the metallicity gradients in the Galactic disk with the tracer population and show that there is detectable azimuthal variation in the metallicity distribution at a given radial annulus. In section~\ref{sec:result}, we also interweave the discussion of our results in the context of what is currently known about the Galactic metallicity gradients along with the how the azimuthal variation might fit into the picture. With new data from the Gaia spacecraft now available, including the spatial positions and chemistry for millions of stars, in section~\ref{sec:future}, we take an initial look at the azimuthal variations in the metallicity structure in the disk with the Gaia DR3 data and disucss future prospects for chemical cartography with this exciting dataset. Finally, we summarize and conclude in section~\ref{sec:summary}

\section{Data} \label{sec:data}
In this section, we describe the data required to carry out aims of this work, namely to explore the metallicity gradient and the azimuthal variation using hot stars from LAMOST and a  complementary data set from Gaia DR3. More specifically, in section~\ref{subsec:selection}, we define the target selection and metallicities derived from LAMOST spectra for the sample.  In section~\ref{subsec:selection}, we discuss the construction of the complementary sample using Gaia DR3. Finally, in section~\ref{subsec:EDR3}, we describe how we determine the spatial positions and velocities for our sample using data from Gaia DR3.

\subsection{The Hot Stars in the LAMOST Survey} \label{subsec:selection}
We begin our investigation with an existing catalogue, presented in \cite{Xiang2021}, of $\sim$332,000 massive (M $\gtrsim$ 1.5~\Msun, e.g., see their Figure~12), hot (\teff\ $>$ 7000~K) stars identified in the LAMOST spectral dataset. The LAMOST survey \citep[see more details at][and references therein]{Cui2012, Zhao2012b} has collected low-resolution (R = $\lambda/\Delta\lambda \sim$ 1800) optical (3800 $< \lambda < $ 9000~\AA) spectra for more than 8 million stars in the northern hemisphere (i.e., declinations larger than --10$^{\circ}$). The spectra have been used to derive stellar atmospheric parameters (i.e., \teff, \logg, \feh, \vsini) and chemical abundances (for up to 16 elements)  through both $\chi^2$ minimization with a grid of synthetic spectra \citep[e.g.,][]{LAMOSTpipeline} as well as through machine learning \citep[e.g.][]{Ting2019, Xiang2019, Xiang2021}. These pipelines have been employed upon mostly FGK-type stars and have enabled the exploration of metallicity gradient of the Milky Way.  Only within the last year, novel machine learning algorithms have been deployed on the $\sim10^5$ hot stars within LAMOST to derive their stellar parameters \citep{Xiang2021}. These newly derived stellar parameters for hot stars represents a large, powerful sample of relatively young stars to not only explore the radial and vertical metallicity gradient but also any azimuthal variations in the metallicity distribution, which has been predicted in recent theoretical work \citep[e.g.,][]{Dimatteo2013, Grand2016, Spitoni2019, Carr2022}. Below we briefly describe the method for selecting and inferring stellar parameters for hot stars in the LAMOST survey.

In order to construct a catalogue of stellar atmospheric parameters, and most importantly metallicity, using LAMOST, \cite{Xiang2021} began by selecting a candidate list of hot stars. They did this by cross matching all stars that are both classified by the LAMOST DR5 and DR6 pipelines as O, B, A and early F-type with an all sky hot star catalogue outlined in \cite{Zari2021}. The latter was based on Gaia DR2 photometric and astrometric data \citep{Gaiasummary2018} along with  complementary data from 2MASS \citep{Cutri2003}. After obtaining $\sim$844,000 candidate hot stars, \cite{Xiang2021} derived the stellar atmospheric parameters (\teff, \logg, \feh, \vsini), which they refer to as stellar labels, using a modified version of The Payne \citep{Ting2019}. This `HotPayne' tool uses a neural network algorithm, which models the fluxes in each spectrum as a non parametric function of the stellar labels and was trained on a synthetic grid of spectra.  For this, a Kurucz model spectral grid was constructed using SYNTHE and convolved to the appropriate resolution with the known line spread function of the LAMOST spectra. A neural network was trained on the large synthetic grid, which enables efficient and precise spectral interpolation. This interpolation is then used to derive the stellar atmospheric parameter with the LAMOST spectrum. We refer the reader to \cite{Ting2019} and  section~4 of \cite{Xiang2021} for a more detailed discussion of The Payne and how it was modified to work with the hot stars covered in this work. While the catalogue presents the atmospheric abundance of silicon (in addition the \teff, \logg, \feh), the authors note that these abundances likely have large systematics and as such we choose not to use these values. 

With these data in hand, we are in a position to determine the radial and vertical metallicity gradient for relatively young hot stars in the Galactic disk. However, before we do that, we must ensure that both the spatial positions and the metallicities for the stars are precise enough to carry out this work. Therefore, we must apply some quality control cuts to the initial dataset to achieve a sample of stars where the spatial positions and metallicities are reasonable. We attempt to limit these cuts to the minimum number required to achieve a large sample while also ensuring precise spatial and chemical properties. There are two main sets of quality criteria cuts: those required to achieve high-quality metallicity from the LAMOST spectra (this section) and those required to achieve high-quality spatial positions (see section~\ref{subsec:EDR3}).

Following the recommendations by \cite{Xiang2021}, we apply the following cuts on the LAMOST hot star catalogue:
\begin{enumerate}
\item Remove stars with \logg\ less than 2~dex or larger than 5~dex -- Stars with \logg\ below 2, which were identified  in section~5.3 of \cite{Xiang2021} to potentially be chemically peculiar stars, likely have incorrectly determined surface gravities. Additionally, we remove stars with \logg\ larger than 5~dex because these stars are either  likely hot subdwarfs where the stellar atmospheric parameters will be poorly determined due to extrapolation. 
\item Remove stars with \vsini\ less than 0 -- Negative \vsini\ values were allowed by their pipeline as an extrapolation and would require a visual inspection of the spectra in order to determine whether it is an artifact or astrophysical. 
\item Remove stars with \teff\ larger than 25 000~K -- These extremely hot O-type stars will likely have significant non-local thermodynamic equilibrium (NLTE) corrections that could cause issues with the reported stellar parameters \citep[e.g., see Figure 11 of][]{Xiang2021}.
\item Remove stars with a signal-to-noise ratio  (SNR) below 30 -- To achieve the high-quality estimates in the stellar atmospheric parameters, a SNR cut had to be applied.
\item Remove stars with `chi2ratio' value larger than 10 -- As noted in section~5.2 of \cite{Xiang2021}, the quality of the spectral fits were defined using a  chi2ratio’ flag. This flag is derived by taking the deviations of the reduced $\chi^2$ from the median value at a given SNR and \teff. By removing stars with a  `chi2ratio' value larger than 10, we are removing stars whose spectral fits have reduced $\chi^2$ values that are 10$\sigma$ outliers compared to others at comparable \teff, and SNR. 
\item Stellar parameters and radial velocities (RVs) must be defined with reasonable uncertainties -- We remove stars with unknown or large uncertainties ($\sigma$RV $>$ 100~\kms) in RV. We do this because ultimately we will want to compute the spatial positions and velocities of our stars and without reasonably precise RVs we will be unable to do this. Further we require that \teff, \logg\ and \feh\ are all known. Since this study will focus on the metallicity gradient in the Galactic disk, it is critical that we have reasonable measured metallicities. As such we further require that star have a metallicity uncertainties, $\sigma$[Fe/H], less than 0.20~dex. We can change this criteria by more than a factor of 2 before seeing any changes in the results. 

\end{enumerate}

In Section 4.7 of \cite{Xiang2021}, they compared their catalogue of stellar atmospheric parameters for hot OBA-type stars against the literature to quantify the level of systematics against other studies.  Their results indicate that stars with \teff\ $>$ 25,000~K or \logg\ $<$ 2~dex, have systematics at the level of 3000~K and 0.80~dex, respectively. This is why we choose to cut stars in these parameter regimes. The authors show (see their Figures 10, 11, and 12) that with the above cuts, there are minimal systematic offsets between the literature (and as a function of SNR) in the atmospheric parameters. We note that while systematic effects are expected to be low, the catalogue employed here from \cite{Xiang2021}, does not  consider NLTE effects, a key limitation to their (and subsequently this) study. 

Upon applying these initial cuts to achieve high-quality chemical properties, we were left with an initial sample of $\sim$140 000 hot stars with \teff\ $>$ 7000~K. We must also  consider quality cuts on the astrometric data required for the spatial positions and namely the derivation of R and Z (see section~\ref{subsec:EDR3}).

\subsection{Complementary Sample: Gaia DR3 Gradient Sample} \label{subsec:DR3RVS}
While the hot star sample from LAMOST allows us to probe a relatively young (<2 Gyr) tracer population, with the release of millions of metallicities all across the sky from Gaia DR3 \citep{Gaiasummary2022, RecioBlanco2022}, we are in a position to have a complementary sample of tracer stars that cover a wider range of azimuthal angles in the Galactic disk. With this complementary sample, we aim to only (1) validate that the methodology we use to derive the radial and vertical gradients give reasonable and consistent result with \cite{GaiaCC22}, and (2) explore the azimuthal variations for stars in the Galactic mid-plane (see section~\ref{sec:future}). We leave further investigations (e.g., exploring the radial and vertical gradients and azimuthal variations as a function of spectral type or age) with this dataset to future studies. 

To construct a complementary Gaia sample, we follow the exact ADQL query from the `gradient sample’ from \cite{GaiaCC22}. This sample is defined in Section 2.5 of \cite{GaiaCC22}. More specifically, to obtain this complementary sample, we begin carrying out the ADQL query  outlined in Listing 3 in Appendix B of \cite{GaiaCC22}. This sample was specifically constructed to optimize the quality of metallicities and spatial positions. To avoid potential selection biases, we choose to specifically focus here on a sample of bright giant stars \citep[e.g.,][]{GaiaCC22}. More specifically, we require that the effective temperature (\teff) measured from the Gaia spectra are less than 4700~K and that the surface gravity was less the \logg\ $<$ 2.0~dex. This is motivated by the desire to have a disk sample of stars that cover a large swath of the disk (hot OBA-type stars are poorly sampled in those stars with quality chemical measurements in Gaia DR3). With the initial gradient sample in hand from the Gaia Archive, we then cross-match that with the distance catalogue of  \cite{Bailer-Jones2021} and remove stars with no distance information. This yielded $\sim$1.03 million stars.  Following \cite{GaiaCC22}, we take the {\sc mh\_gspspec} column in the Gaia DR3 {\sc astrophysical\_parameters} table as our Gaia DR3 metallicities. These metallicities are derived using spectra from the moderate resolution (R$\sim$11 500), optical (8450 -- 8720~\AA) Gaia Radial Velocity Spectrograph (RVS) instrument \citep[we refer the reader to][for more details on the RVS spectra and their analysis]{RecioBlanco2022}. The median uncertainty in metallicities  in the gradient sample catalogue of giant stars is 0.04~dex. With a high-quality chemical sample from Gaia DR3, we must apply a set of quality cuts on the astrometric data required for the spatial positions and namely the derivation of R and Z.

\subsection{Galactic Positions and Kinematics with Gaia DR3} \label{subsec:EDR3}
In order to obtain photometric and astrometric data for our sample we made use of data release 3 (DR3) from the Gaia spacecraft \citep{Gaiasummary2022}. We cross matched the $\sim$ 140 000  hot stars with quality stellar atmospheric parameters (see section~\ref{subsec:selection} for more details) with Gaia DR3 using a 2\arcsec search radius. While stellar distances, required to determine the spatial position of each star, can be derived as the inverse of the parallax measured by Gaia DR3, parallax inversion can lead to significantly biased distances \citep[e.g.][]{Astraatmadja2016}. Therefore, distances for each star were sourced from the catalogue outlined in \cite{Bailer-Jones2021}. More specifically, we adopted the geometric distance estimates. This choice was motivated by the fact that we do not want to rely on color and magnitude information in the distance inference and as such it is recommended in section~5.3 of \cite{Bailer-Jones2021} to favor the geometric over photo-geometric distance estimates.  In short, these distances are derived in a probabilistic Bayesian framework, whereby the posterior probability of a `true' distance to each star given a measured parallax and its associated uncertainty is determined (see their Equation 1). The method adopted by these authors require a distance prior, a term which encodes the prior information that we have about the distances of stars in the Galaxy. The authors adopt a prior that is constructed from a model of the Milky Way (see their Equation 3 and their Figure 2 for more information on the distance prior). While, in principal, the prior affects the posterior distribution, in the limit where the uncertainties on the data (in this case the parallaxes reported from Gaia) are high-quality, the prior will not affect the posterior significantly. 

For the purposes of this study, we want fairly precise spatial positions so that we can quantify the vertical and radial metallicity gradient. As such we limit ourself to stars which have parallax uncertainties better than 30\% as well as remove stars with inferred distances from \cite{Bailer-Jones2021} with uncertainties larger than 30\%. Together these cuts along with those outlined in section~\ref{subsec:selection} reduces the initial LAMOST sample to $\sim$135 000 stars. Additionally, this reduces the initial Gaia gradient complementary sample to $\sim$840 000 giant stars. The median parallax uncertainty in both samples is around $\sim$3\%.

With distances for each star, we compute the full 6 dimensional spatial position and velocity vectors for each star and its associated uncertainty. The spatial positions (X, Y, Z location) and velocities (U, V, W) in a 3-dimensional Galactocentric-Cartesian frame were determined using Astropy's \code{SkyCoord} class \citep{Astropy} with Gaia DR3 right ascension and declination, proper motions, with distances adopted from \citet{Bailer-Jones2021}, and RVs adopted from LAMOST for the hot stars. RVs were adopted from Gaia DR3 for the complementary sample.   We assume that the solar position  is \{X, Y, Z\}$_{\sun}$=  \{8.27, 0.00, 0.029\}~kpc and the Sun's velocity in this reference frame is   \{U, V, W\}$_{\sun}$=  \{-12.9, 245, 7.78\}~\kms \citep{Drimmel2018, Gravitycollab2021}. Uncertainties in the 6 dimension phase space (X, Y, Z, U, V, W) were determined by a Monte Carlo simulation with 1000 realizations, where each observational variable (i.e. right ascension and declination, proper motion, RV, and distance) were sampled from a random normal distribution. 

\begin{figure}
	 \includegraphics[width=1\columnwidth]{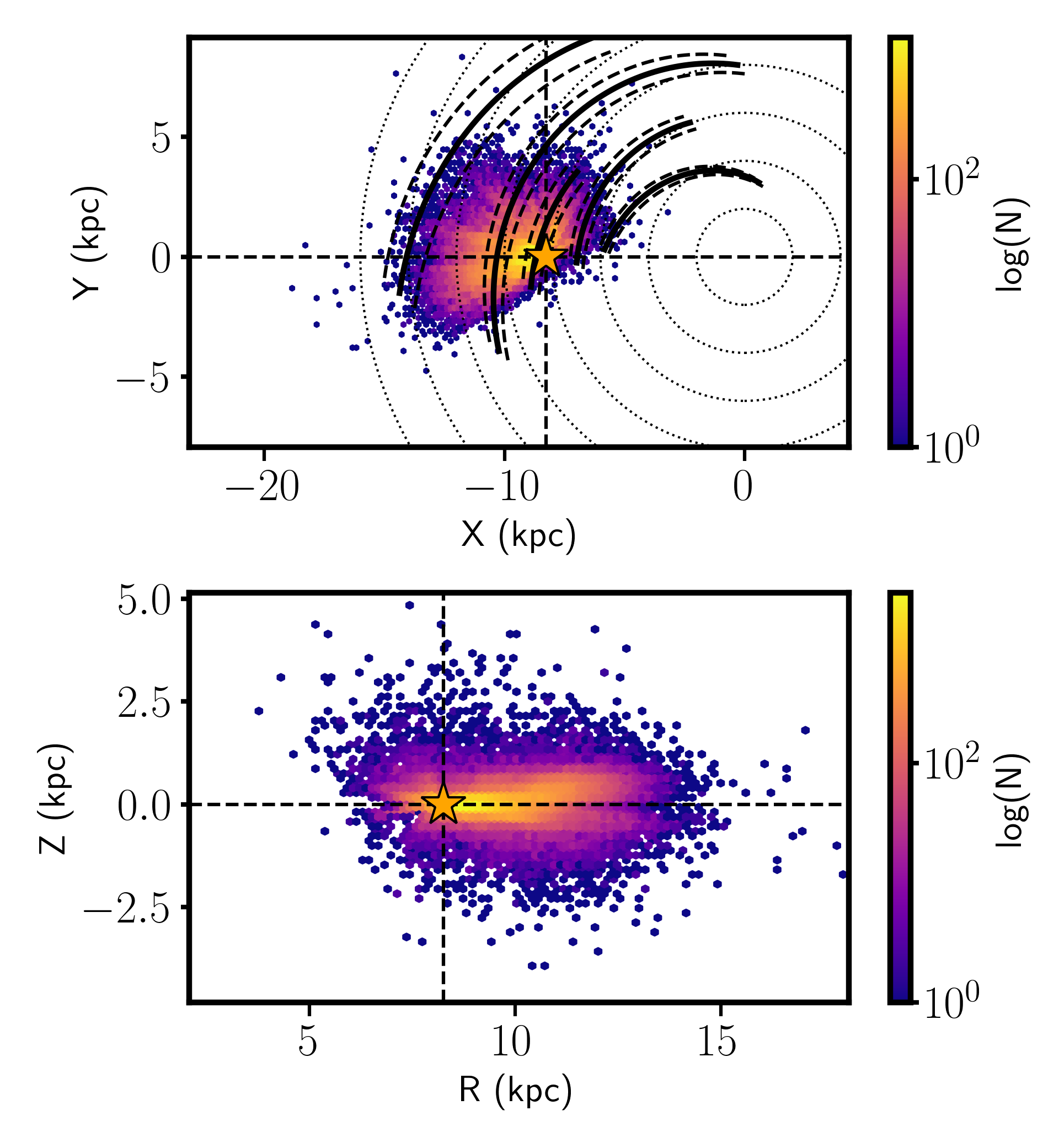}
	\caption{Top: Two dimensional density diagram of the X-Y spatial positions of the hot OBAF star sample in LAMOST. For reference, the solar position is denoted as an orange star. The approximate location of the spiral arms in the Galaxy from \protect\cite{Reid2014} are shown as thick black lines and their extend is marked by thick dashed lines. The color coding represents the number density of stars in log(N). Bottom: Two dimensional density diagram of the R-Z spatial positions of our full sample. The color coding represents the number density of stars in log(N). } 

	\label{fig:XYRZ_N}
\end{figure}

We know from previous studies \citep[e.g.][and references therein]{Bilir2012, Boeche2013b, Boeche2014, Onal2016, Yan2019} that metallicity gradients can and do depend upon whether one samples stars from the (kinematic) thin or thick disk. Therefore, we made use of the velocities derived above to probabilistically separate our stars into a (kinematic) thick and thin disk subpopulation. We computed the probability of a star belonging to the (kinematic) thin disk, thick disk, and stellar halo using the method outlined in \cite{Ramirez2013}. With these probabilistic memberships, in both samples we remove probable halo stars (i.e., stars that have a 1\% or higher chance of belonging to the stellar halo) since in this study we are focused on the Galactic disk. With this cut our final sample of \numstar\ OBAF-type stars (and 847 214 complementary giant stars from Gaia DR3), which we will use to explore the radial and vertical metallicity gradient. The typical (median) uncertainties are 85~K, 0.10~dex, 0.08~dex in \teff, \logg, and \feh, respectively for the host star catalogue. The typical (median) uncertainties are 30~K, 0.10~dex, 0.04~dex in \teff, \logg, and \feh, respectively for the complementary Gaia DR3 sample. We note here that nearly all ($>$95\%) of the sample is high probability ($>$90\%) thin disk stars and as such with this particular sample is well suited for the exploration of the metal gradients in the thin disk. 

\begin{figure}
	 \includegraphics[width=1\columnwidth]{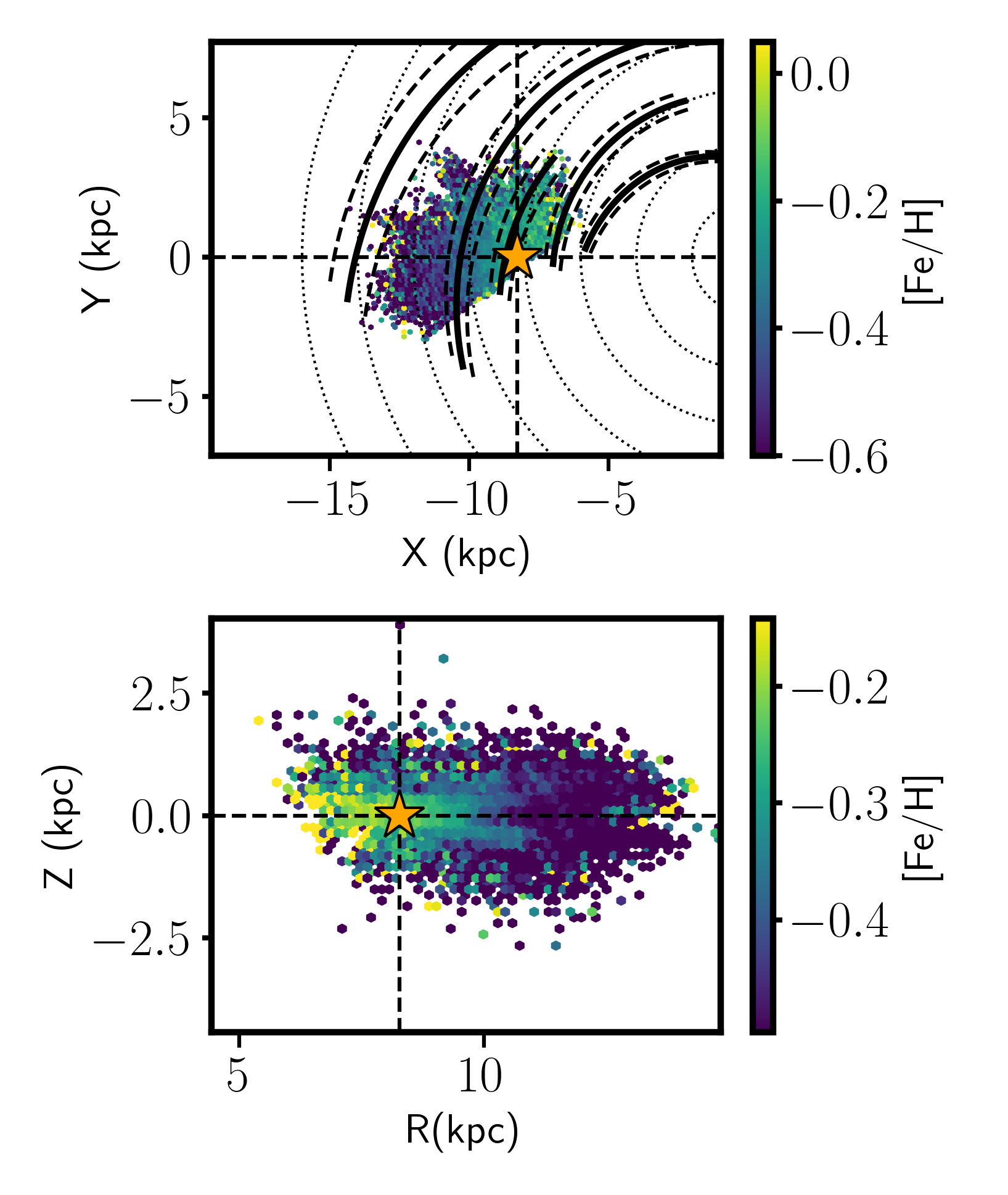}
	\caption{Top: The X-Y spatial distribution of \feh\ for the LAMOST sample. The color coding represents median [Fe/H] in each hexagonal bin. Each hexagonal bin were selected to have a minimum of 5 stars.  The reference background is the same as in Figure~\ref{fig:XYRZ_N}. Bottom:  R-Z spatial distribution of \feh\ across the Galactic disk. Similar to the top panel, the color coding represents median [Fe/H] in each hexagonal bin. It is clear from the top panel that the inner Galaxy is more metal rich compared to the outer Galaxy (i.e., negative radial metallicity gradient). The bottom panel illustrates that the stars in the plane of the Galaxy are more metal enhanced compare to stars which are at larger vertical distance (i.e., negative vertical metallicity gradient).   }   
	\label{fig:XYRZ_FEH}
\end{figure}

In Figure~\ref{fig:XYRZ_N}, we show the X-Z (top) and R-Z (bottom) spatial distribution of the final sample of \numstar\ OBAF-type stars. The color coding in each hexagonal bin represents the logarithm of the number of stars (log(N)) located in that bin. For reference, we also show the location of the Sun in both panels as a orange star as well as the approximate location of the spiral arms \citep[taken from ][and shown as thick bold black lines surrounded by thick dashed lines]{Reid2014}. We can see that the sample covers Galactocentric radii between $\sim$6.8--12~kpc and have a vertical extent of |Z| $\leq$ 2~kpc. In Figure~\ref{fig:XYRZ_FEH}, we make the same plot as in Figure~\ref{fig:XYRZ_N}, but instead of color coding by the log(N), each hexagonal bin is colored by the median metallicity. The top panel of Figure~\ref{fig:XYRZ_FEH}, shows that the inner Galaxy has, on average, a larger metallicity compare to the outer Galaxy (i.e, the hot stars in LAMOST show a negative metallicity gradient). Additionally, the bottom panel of Figure~\ref{fig:XYRZ_FEH}, shows that the metallicity decreases as |Z| increases (i.e. a negative vertical metallicity gradient). In the following sections we will aim to quantify these qualitative trends.

\section{Methods: Deriving the Metallicity Gradient and Azimuthal Structure} \label{sec:Methods}

With the spatial and chemical information for \numstar\ hot, massive stars spread out across the Galaxy, we are now in a position to derive the vertical and radial metallicity gradients. In addition, we will also constrain any azimuthal variations in the radial gradient. In this section, we will outline the methods used to derive the vertical and radial metallicity gradients and the azimuthal variations.
 
We choose to model both the radial and vertical metallicity gradients as linear functions with an intrinsic scatter. That is to say, that we model the radial metallicity profile of the Galaxy, as
\begin{equation}
\mathrm{[Fe/H]_R }=  \frac{\Delta \mathrm{[Fe/H]}}{\Delta \mathrm{R}} R   + b_{R}
\label{eq:Mgrad}
\end{equation}
\noindent In Equation~\ref{eq:Mgrad}, the metallicity at a particular Galactocentric radius (R) in the Galaxy, $\mathrm{[Fe/H]_R }$,  is linear function of the Galactocentric radius of a star (R) and the radial metallicity gradient, is denoted as \Mgrad. Additionally, $b_{\mathrm{R}}$ represents the metallicity of the Galaxy at the Galactic center (i.e., R=0) if this metallicity gradient were valid across the full disk. While it is common in the literature to model the radial metallicity gradient as a simple linear function, it has been shown in young open cluster systems that the metallicity gradient is different in the outer disk compared to the inner disk with a break at Galactocentric radii of $\sim$12-13 kpc \citep[e.g.,][]{Spina2022}. Since our sample of OBAF stars from LAMOST have a median R = 9.3 $\pm$ 1.15~kpc, nearly all of our data is encompassed between 6 $<$ R $<$ 12~kpc. As such, we choose a single linear model rather than a broken piecewise linear function.

We choose to model the vertical metallicity gradients as linear function such that,
\begin{equation}
\mathrm{[Fe/H]_Z}=  \frac{\Delta \mathrm{[Fe/H]}}{\Delta \mathrm{Z}} \mathrm{|Z|}   + b_{Z}
\label{eq:Vgrad}
\end{equation}

\noindent where in Equation~\ref{eq:Vgrad}, the metallicity at a particular absolute vertical height away from the Galactic plane, $\mathrm{[Fe/H]_Z }$,  is linear function of its absolute vertical height and the vertical metallicity gradient is denoted as \Vgrad.  In this case, $b_{Z}$ represents the metallicity of stars that are in the mid-plane of the Milky Way (i.e., with |Z| = 0~kpc). In both cases, we will also fit an intrinsic scatter term ($\boldsymbol{\Lambda_R, \Lambda_Z}$)\footnote{We will denote varibles that matrix as bold upper case symbols.}, which represents the intrinsic dispersion around the radial and vertical metallicity gradient, respectively. We note that $\boldsymbol{\Lambda_R}$ and $\boldsymbol{\Lambda_Z}$ are actually 2$\times$2 tensors, which not only contain the dispersion around the line (denoted by $\lambda^2$) but also define the direction of the scatter (i.e., we assume that $\boldsymbol{\Lambda_R} =\frac{\lambda_R^2}{1+m_R^2} \big(\begin{smallmatrix}
(\Delta \mathrm{[Fe/H]}/\Delta R)^2 & -\Delta \mathrm{[Fe/H]}/\Delta R
\\
-\Delta \mathrm{[Fe/H]}/\Delta R
 & 1
\end{smallmatrix}\big)$). This assumption is essentially that the scatter is perpendicular to the linear function with some variance ($\lambda_R^2$).  Some limitations to this choice is that, we assume that the intrinsic scatter in the metallicity gradients are independent of location and that it is perpendicular to the linear relation between metallicity and radius (and height above/below the plane), which may not necessarily be true. We have also tried a model with intrinsic scatter that is only along the \feh\ axis and found identical results in the gradients. Now that we have written down how we will model the data, we must choose a method to fit our model and solve for the unknown parameters (i.e.,\Mgrad, $b_R$, \Vgrad, $b_Z$, $\lambda_R$, $\lambda_Z$). In order to fit these linear functions, we follow the steps outlined in Section~7 and 8 of \cite{Hogg2010}\footnote{A python implementation of fitting a line to data including both intrinsic  scatter and uncertainties in two dimensions can be found at \url{https://dfm.io/posts/fitting-a-plane/}.} . Namely we adopt a Bayesian approach because we can readily fit our model while accounting for uncertainties in R (or Z) and \feh, simultaneously. Under this approach, we aim to take our noisy measurements of R, Z, and \feh\ and use them to infer the parameters of our linear model. Since we are more carefully modeling the uncertainties in the observable parameters under this Bayesian Formalism \citep[e.g.,][]{Anders2017}  we can more readily and accurately determine the uncertainties on our model parameters (i.e., the vertical and radial metallicity gradients).

To formulate this concept more generally, let's say we have made noisy measurements, \{$y_i, x_i$\} with uncertainties described by the 2$\times$2 covariance matrix, $ \boldsymbol{S_i}= \big(\begin{smallmatrix}
\sigma_{x,i}^2 & \sigma_{xy,i}\\
\sigma_{yx,i} & \sigma_{y,i}^2
\end{smallmatrix}\big)$\footnote{In this pedagogical example, the noisy measurements, $y_i$ and $x_i$, could be thought of as the derived metallicity and Galactocentric radius (or absolute vertical height above the disk) to the $i$th star in our dataset, respectively. Therefore, $\boldsymbol{S_i}$  represents the covariance matrix that describes uncertainty in \feh\ and R (or Z) and any covariances between them.}. We want to model  \{$y_i$\}  as a linear function of  \{$x_i$\}  with some slope ($m$), intercept ($b$), and intrinsic scatter around this linear relationship ($ \boldsymbol{\Lambda}$). Under a Bayesian inference framework, we must start with with Bayes' theorem, which states that: 

\begin{equation}
p(m, b, \boldsymbol{\Lambda} | y_i,x_i,\boldsymbol{S_i} ) \propto p( y_i, x_i, \boldsymbol{S_i} | m, b, \boldsymbol{\Lambda} ) p(m, b,\boldsymbol{\Lambda})
\label{eq:baye}
\end{equation}
 
\noindent where in Equation~\ref{eq:baye}, $p(m, b, \boldsymbol{\Lambda} | y_i,x_i,\boldsymbol{S_i} )$ is the posterior probability distribution of the model parameters ($m, b, \boldsymbol{\Lambda} $) given the noisy measurements ($y_i,x_i, \boldsymbol{S_i} $), $p( y_i, x_i, \boldsymbol{S_i} | m, b, \boldsymbol{\Lambda} )$ is the likelihood of obtaining the noisy measurements given a defined model, and $p(m, b,\boldsymbol{\Lambda} )$ represents the prior information we have of the model parameters. In our case, the marginalized likelihood can be written down as:

\begin{equation}
p( y_i, x_i, \boldsymbol{S_i} | m, b, \boldsymbol{\Lambda} )  \propto \prod_{i=1}^{N}\frac{1}{\sqrt{2\pi\Sigma_i^2}}\exp{\left ( -\frac{\Delta_i^2 }{2\Sigma_i^2}  \right )}
\label{eq:like}
\end{equation}

\noindent In Equation~\ref{eq:like}, we define a Gaussian likelihood function where $\Delta_i^2  = y_i - (mx_i -b)$ and represents the difference between the measured and predicted value of $y$ based on the linear model  for the $i$th data point. Additionally, $\Sigma_i^2 =  \boldsymbol{v}^T(\boldsymbol{S_i} + \boldsymbol{\Lambda})\boldsymbol{v} $, where here $\boldsymbol{v}^T$ equals the 1$\times$2  matrix ($-m$ 1). We encourage the reader to consult \cite{Hogg2010} and references therein for the detailed derivation of this marginalized likelihood probability distribution. In addition to the likelihood function, which is written in Equation~\ref{eq:like}, we also need to specify any prior information that we have about the model parameters, i.e.,  $p(m, b, \boldsymbol{\Lambda})$. We choose relatilvey uninformative priors. Specifically, we select a uniform probability distribution in $m$ between -2 and 2~dex/kpc, a uniform prior in $b$ between -2 and 2~dex and a uniform prior in $\lambda$ between 0 and 1~dex. 

Ultimately, in Bayesian inference, we want to learn the full posterior distribution of the model parameters and how it depends on the observed data. To learn the model parameters and their uncertainties, we explore the multi-dimensional parameter space and find  where the conditional probability of the model given the data is maximized. Since we do not know, a priori, the model parameters,  we cannot solve this conditional probability. As such, maximizing the posterior probability requires maximizing the likelihood function multiplied by the prior (both of which we know). We use a Markov Chain Monte Carlo (MCMC) sampler set up with 10 walkers and 1000 steps each to explore the parameters space. Specifically, we use the python package \code{emcee} \citep{emcee} to sample the parameters space and use the chains to derive the model parameters and their uncertainties. We use a rigorous Bayesian fitting procedure here to better account for the uncertainties in the data \citep[similar Bayesian fitting procedures like the one outlined above can also be found in the literature, e.g.,][]{Anders2017}.

Once we fit the model, we can plot the residuals of the data subtracted from the model as a function of X-Y spatial position and search for azimuthal structure in the residuals.  If the Galactic disk has the same (linear) radial metallicity gradient along all azimuthal angles, there should be no azimuthal structure in the residuals between our predicted metallicity and the observed ones. However, if the radial metallicity gradient changes with azimuthal angle, we would find structure in the residuals that would be spatially/azimuthally dependent. Recent studies have suggested there could be azimuthal variations at the 0.05--0.10~dex level in the metallicity gradient due to interactions with dwarf galaxies \citep[e.g.,][]{Carr2022}, or a result spiral arm density fluctuations and other secular processes \citep[e.g.,][]{Dimatteo2013, Khoperskov2018,Fragkoudi2018, Spitoni2019, Wheeler2021,Filion2023}. Part of our aim is to observationally test that hypothesis. 

In following section (Section~\ref{sec:result}), we apply our methods outlined above to the metallicities and spatial positions of \numstar\ hot LAMOST stars discussed in section~\ref{sec:data} to derive the vertical and radial metallicity gradients in the Galactic disk. We also applied the above radial gradient methodology to the Gaia DR3 gradient sample (see section~\ref{subsec:DR3RVS}) and have validated that we can recover the radial and vertical gradients from \cite{GaiaCC22} using this sample and the above methodology. 
\section{Results and Discussion} \label{sec:result}

\subsection{Radial Metallicity Gradient} \label{subsec:radial}

We begin by quantifying the radial metallicity gradient in the Galactic disk with the sample of \numstar\ OBAF-type stars observed in the LAMOST survey. It is clear even in Figure~\ref{fig:XYRZ_FEH}, that the inner Galaxy (with R$<$~8.3~kpc) is more metal rich, on average compared to the outer Galaxy. It is also clear from that figure \citep[as well as previous studies of the Galactic metallicity gradient,][and references therein]{Onal2016, Anders2017, Yan2019}, that the value of radial metallicity gradient is dependent upon vertical extent of the sample. For this reason, we split our dataset into equally-sized vertical bins ranging from  0 $<$ |Z| $<$ 1.5~kpc with a bin width of 0.2~kpc. We also require a minimum of 500 stars in each bin in order to adequately measure the radial metallicity gradient. For each bin in |Z|, we derive the parameters of the linear function ($m_R, b_R, \lambda_R$) defined in Equation~\ref{eq:Mgrad} using the method outlined in section~\ref{sec:Methods}. 

\begin{figure}
	 \includegraphics[width=1\columnwidth]{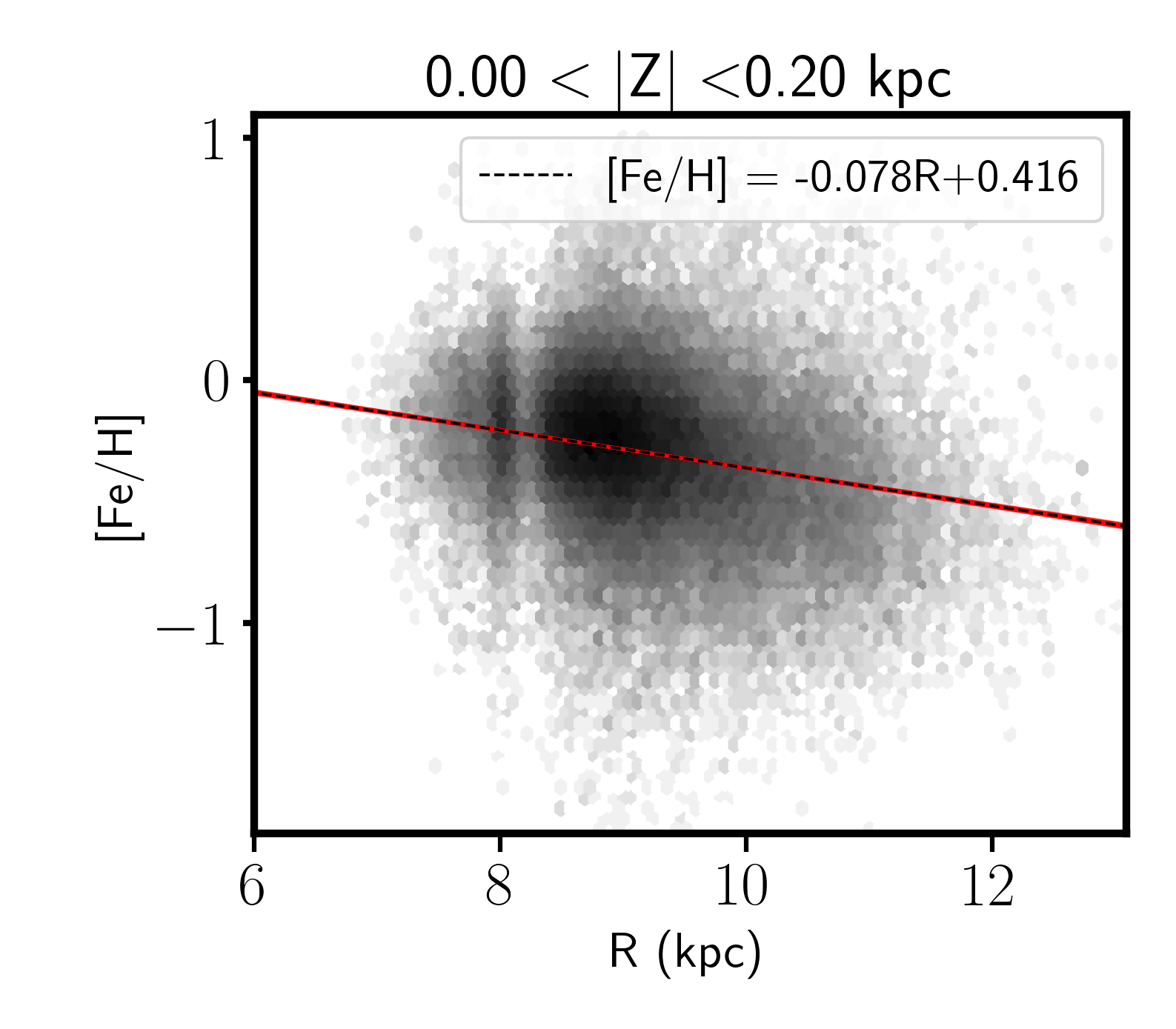}
	\caption{The \feh\ as a function of Galactocentric radius (R) for OBAF-type stars in LAMOST with 0.0 $<$ |Z| $<$ 0.2~kpc (grey-scaled background). The dashed black line represents the best fit linear model determined using Bayesian inference linear regression. Also shown are thin red lines (which make a red band due to highly constrained model) of 100 randomly selected linear models realizations from the Bayesian MCMC sampling. The derived metallicity gradient for these stars is \Mgrad\ = --0.078 $\pm$ 0.001~dex/kpc. }   
	\label{fig:Mgrad_data}
\end{figure}
 
As an illustrative example of this fitting procedure, in Figure~\ref{fig:Mgrad_data}, we plot the \feh\ as a function of Galactocentric radius (R) for stars in our final hot star sample with vertical heights above (or below) the disk mid-plane of |Z| $<$ 0.20~kpc. The \feh\ and R for these stars (background grey density), along with their respective uncertainties, are then input into our Bayesian inference (described in section~\ref{sec:Methods}).  The solid thick red band in Figure~\ref{fig:Mgrad_data} is the best fit linear model from the Bayesian linear regression along with thin red lines, which represent 100 draws from the posterior. We find a negative metallicity gradient of \Mgrad\ = --0.078 $\pm$ 0.001~dex/kpc for OBAF-type stars with 0.0 $<$ |Z| $<$ 0.2~kpc. We then apply the above procedure to all bins in |Z| and tabulate the results of our model fitting for each bin in Table~\ref{tab:Mgrad}.

\begin{figure}
	 \includegraphics[width=1\columnwidth]{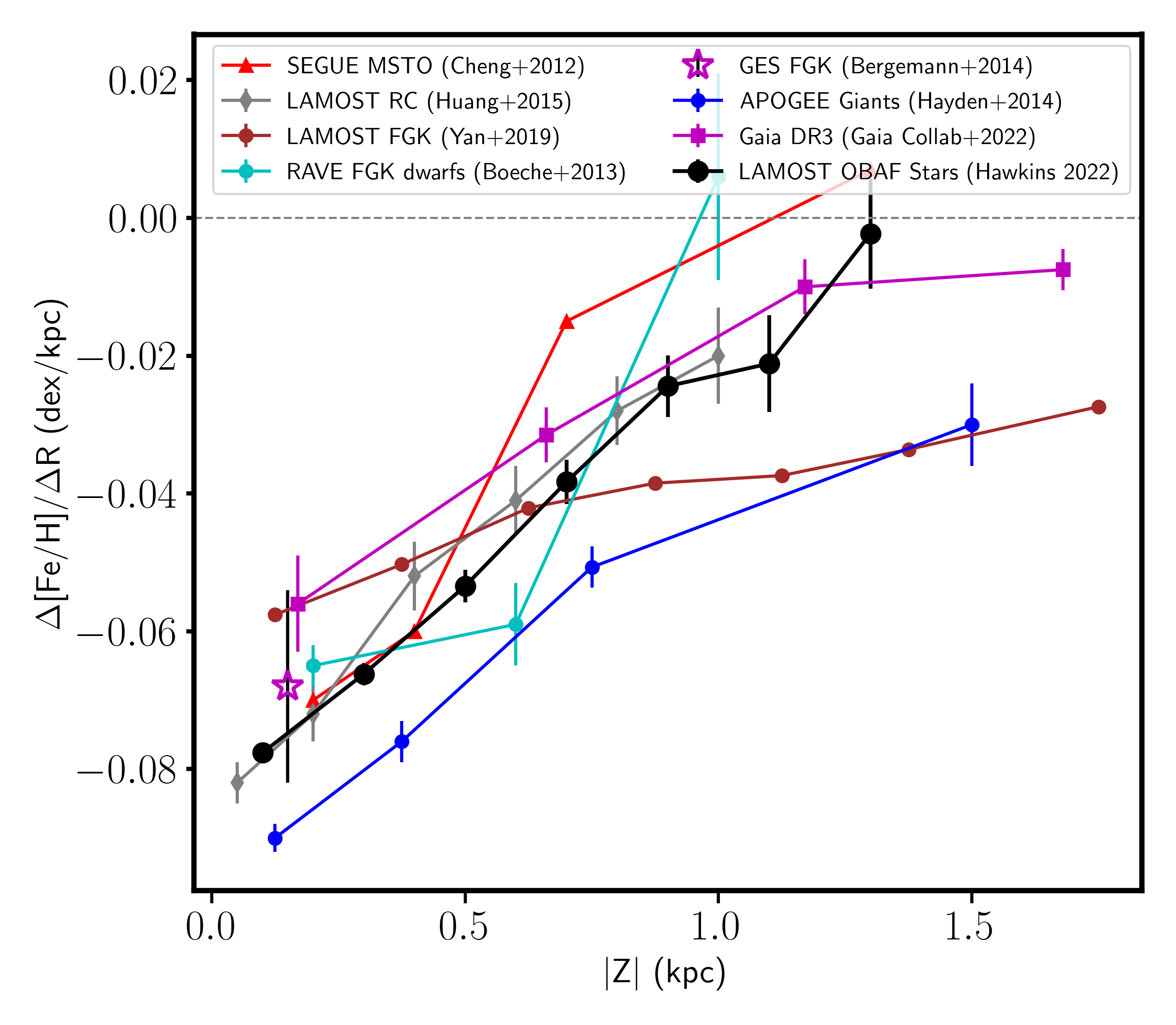}
	\caption{The metallicity gradient, \Mgrad, as a function of absolute vertical height (|Z|) away the Galactic mid-plane for for OBAF-type stars (black circles/line) used in this study. For comparison, we also show the radial metallicity gradient measured using: main sequence turn off stars in SEUGE \protect\citep[][red triangles/line]{Cheng2012},  LAMOST survey red clump type stars  \protect\citep[][grey diamonds/line]{Huang2015},  LAMOST survey FGK-type stars \protect\citep[][brown circles/line]{Yan2019}, RAVE survey FGK dwarf stars \protect\citep[][cyan cirlces/line]{Boeche2013}, Gaia-ESO survey FGK-type stars \protect\citep[][magenta star]{Bergemann2014}, APOGEE red giant branch stars  \protect\citep[][blue circles/line]{Hayden2014}, and FGK dwarfs and giants from the Gaia DR3 survey \protect\citep[][magenta squares/line]{GaiaCC22}. The Galactic radial metallicity gradient becomes shallower (i.e., \Mgrad\ becomes less negative and closer to zero) with increasing height above the mid-plane. The radial metallicity measured using LAMOST OBAF-type stars (black line) is also consistent with other tracers (colored lines).  }   
	\label{fig:MgradZ}
\end{figure}

\setlength{\tabcolsep}{2pt}
\begin{table}
\caption{Radial Metallicity Gradient Fit Parameters for |Z| bins}
\begin{tabular}{llllllll}
\hline\hline
|Z| & \Mgrad & $\sigma$\Mgrad & $b_R$ & $\sigma b_R$ & $\lambda_R$ & $\sigma\lambda_R$ & N\\
(kpc) & (dex/kpc) & (dex/kpc) & (dex)& (dex) & (dex) & (dex)& \\
\hline
0.10 & -0.078 & 0.001 & 0.418 & 0.012 & 0.319 & 0.001 & 77997 \\
0.30 & -0.066 & 0.002 & 0.294 & 0.016 & 0.321 & 0.001 & 30653 \\
0.50 & -0.053 & 0.002 & 0.135 & 0.024 & 0.321 & 0.002 & 11480 \\
0.70 & -0.038 & 0.003 & -0.043 & 0.033 & 0.314 & 0.003 & 4948 \\
0.90 & -0.026 & 0.005 & -0.211 & 0.049 & 0.322 & 0.006 & 1897 \\
1.10 & -0.022 & 0.007 & -0.280 & 0.071 & 0.338 & 0.009 & 830 \\
1.30 & -0.001 & 0.008 & -0.488 & 0.084 & 0.330 & 0.011 & 506 \\
\hline\hline
\end{tabular}
NOTE: Column 1 is center of the vertical |Z| bin.  The derived radial metallicity gradient slope and its uncertainty can be found in columns 2 and 3, respectively. The intercept of the linear model and its uncertainty are tabulated in columns 4 and 5, respectively. The intrinsic scatter around the linear radial metallicity gradient relation and its uncertainty can be found in columns 6 and 7, respectively. 
\label{tab:Mgrad}
\end{table}

In Figure~\ref{fig:MgradZ}, we plot the value of the radial metallicity gradient, \Mgrad, as a function of |Z| for all bins where the number of stars in the bin is larger than 500. For reference, we also plot the \Mgrad\ from a representative (but very incomplete) set of previous studies from different surveys with a variety of tracers \citep{Cheng2012, Boeche2013, Hayden2014, Bergemann2014, Huang2015}. These were selected to range across different tracers populations and different observational surveys. Figure~\ref{fig:MgradZ}, illustrates that Galactic disk has negative (--0.10 $<$ \Mgrad\ $< 0.00$~dex/kpc) and flattening radial metallicity gradient with increasing absolute vertical height above (or below) the Galactic mid-plane. Quantitatively, we  find that the value of the metallicity gradient that we measure with the OBAF-type tracers from LAMOST in this work are broadly consistent with previous literature using a wide variety of stellar tracers \citep[e.g., See Table 1 of][and references therein]{Onal2016}. We note that the radial metallicity gradient that we find is generally shallower, by $\sim$0.015~dex/kpc, across all vertical heights compared to the measured value by \cite{Hayden2014} using APOGEE red giant stars as a tracer. However, the results from that study tend to show steeper radial metallicity gradients than what is found in other studies of similar type stars \citep[e.g., see Table 1 of][and references therein for a sense of the variation of radial metallicity gradient values as measured by different tracers]{Onal2016}. It is evident in previous studies, that the variation in the derived radial and vertical metallicity gradient  may be a result on the spatial (vertical or radial) extent of the stellar population being probed.

In addition to illustrating that the radial metallicity gradient becomes shallower at higher vertical heights above (or below) the Galactic disk, we also find that the intercept ($b_R$) of the linear model (i.e., the metallicity of stars near the Galactic center assuming the radial metallicity gradient is valid across the full disk) decreases with increasing vertical distance away from the Galactic mid-plane (i.e., see Table~\ref{tab:Mgrad}). We will discuss this vertical metallicity gradient more in section~\ref{subsec:vertical}.

Finally, we find that the intrinsic scatter about the radial metallicity gradient is roughly constant across all vertical heights and is $\sim$0.30~dex.  Since few works in the literature fit for the intrinsic scatter it is difficult to compare this with previous studies. However,  \cite{Anders2017} applied a similar methodology to the one outlined in section~\ref{sec:Methods}). In that work, the authors modeled the metallicity (and its scatter) of red giants observed in APOGEE as a linear function of their Galactocentric radii accounting for the age of the population. For young stars (i.e. those with ages $<$ 2~Gyr), \cite{Anders2017} found that the radial metallicity gradient was --0.066 $\pm$ 0.007~dex/kpc and became shallower with increasing age. This value is consistent with the our results for stars that are in the Galactic mid-plane. They also found that the scatter around the radial metallicity gradient was  a constant of $\sim$0.15~dex regardless of spatial position of the sample.  While we found a larger scatter ($\sim$0.30~dex) we found that this roughly constant across all vertical heights and radii, consistent with \cite{Anders2017}. The larger scatter we find could be a result of underestimated metallicity uncertainties in the LAMOST dataset. Taken together, these results suggests that the inner Galaxy (in the ranges probed by this study), while more metal-rich than the outer disk, becomes chemically more similar compared to the outer Galaxy as one goes away from the mid-plane. This could be a result of several physical processes, such as radial mixing \citep[e.g.][]{Hayden2015, Anders2017}.



\subsection{Azimuthal Variations} \label{subsec:azimuthal}
\begin{figure*}
	 \includegraphics[width=2.1\columnwidth]{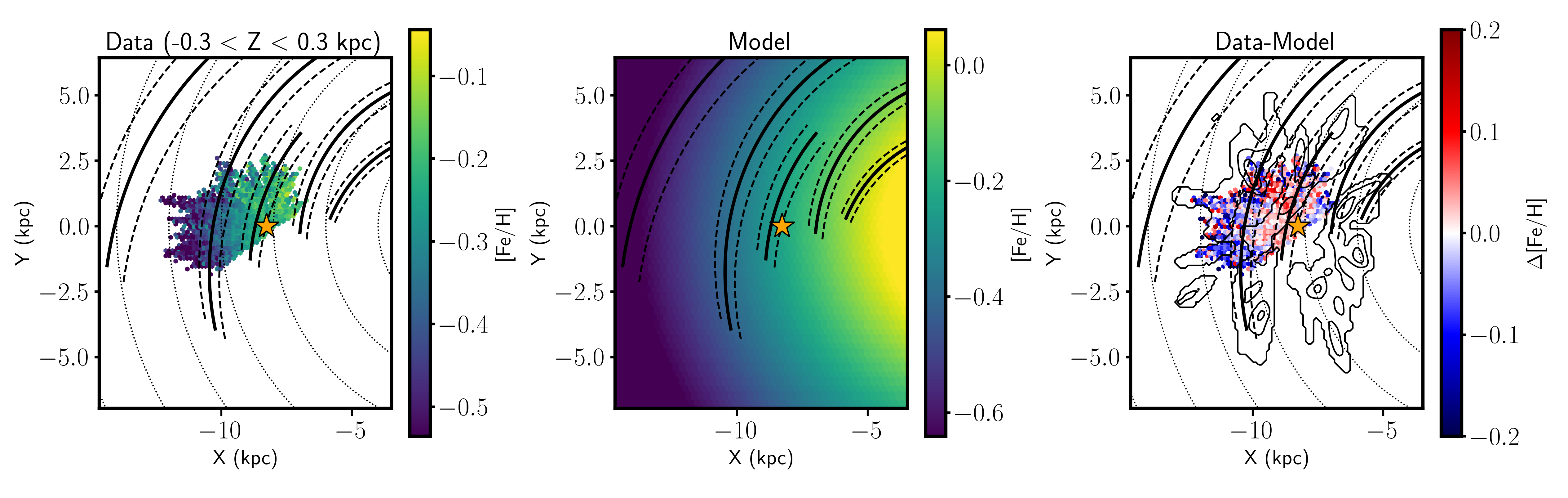}
	\caption{Left: The X-Y projection for the $\sim$97 000 stars with 0.00 $<$ |Z| $<$ 0.30~kpc color-coded by the median metallicity derived with the LAMOST spectra. Middle: The X-Y projection of the metallicity gradient defined by the best fitting model.  Right: The X-Y projection of the residual (color bar) between the observed metallicity (right panel) and the predicted metallicity (middle panel). For reference, in all panels the position of the spiral arms (as determined by high mass star forming regions) from \protect\cite{Reid2014} are shown as black lines (both dashed and solid). In the right panel, we also show the position of the spiral arms (black contour), as determined by the upper main sequence from \protect\citep{Poggio2021}. The residual, $\Delta$\feh, is defined as the observed metallicity subtracted from the predicted metallicity of each star given its Galactocentric radius. We find that there is structure at the 0.05--0.20~dex level. These residuals can be as large as 3 times the typical uncertainties in \feh. } 
	\label{fig:resid}
\end{figure*}
In addition to deriving the radial metallicity gradient (see section~\ref{subsec:radial}), in this study we are also interested in whether there are azimuthal variations in the metallicity distribution (or radial gradient). Such azimuthal variations in the metallicity gradient (and thus the metallicity distribution across the disk) have been predicted in simulations \citep[e.g.][]{Dimatteo2013, Grand2016, Khoperskov2018, Fragkoudi2018, Spitoni2019, Wheeler2021, Carr2022}. Additionally, \cite{Hwang2019}, uncovered azimuthal variations in the metallicity distribution of external galaxies observed in the MANGA survey (e.g., see their Figure 13). Interestingly, azimuthal variation were detected in close pair systems where a host galaxy had a companion nearby. While, small azimuthal variations (at the 0.02~dex level) have been detected in the Milky Way disk on relatively local samples of stars \citep[e.g.][]{Bovy2014, Antoja2017}, in this work we expand to well beyond the solar neighborhood and use a different tracer population. 

To quantify whether there are azimuthal variations, in Figure~\ref{fig:resid}  we show the median metallicity as a function of X-Y position for stars with |Z| $<$ 0.30~kpc (left panel). In addition, we also show the best-fitting  model (using the procedure outlined in section~\ref{sec:Methods}) of the expected metallicity in the same X-Y space (middle panel). The the residual between the observed metallicity at a given X-Y cell and the predicted metallicity in that same cell based on the radial metallicity relation is shown in the right panel. For reference, in all panels we show the location of the spiral arms as identified using high-mass star-forming regions \citep[shown as black lines,][]{Reid2014}. In the right panel, we also show the location of the spiral arms, as determined by the over density of the upper main sequence stars in Gaia \citep[shown as black contours][]{Poggio2021}. Critically, if the same linear radial gradient model were a good fit to the data at all azimuthal angles (i.e., little to no azimuthal metallicity structure), the residuals should be close to zero for the full X-Y extent of the data (and along all azimuthal angles).

The right panel of Figure~\ref{fig:resid} illustrates that we find deviations upwards of 0.20~dex (at the $\sim$2--3$\sigma$ level) between the predicted and observed metallicity at a give X-Y position in the Galactic disk. Additionally, the residual level (indicated by the color bar in the right panel of Figure~\ref{fig:resid}) appears to be dependent on where in the Galactic disk the stars are located. For example, we find that the observed metallicities at around X $\sim$ --8.6~kpc and Y $\sim$ 1.5~kpc are higher than what is predicted from a our simple linear radial metallicity model. We also find that the observed metallicities at X $\sim$ --10.5~kpc and Y $\sim$ -1.5~kpc are, on average, lower than what is predicted from our linear model. We also see hints that the azimuthal variations are largest in the outer rather than inner Galaxy (in the range probed by this study), though a larger dataset that encompasses a wider swath of the Galaxy is required. Interestingly, for this OBAF-type sample of stars from LAMOST, {\it we find that the azimuthal variations does not follow the expected location of the spiral arms} \citep[unlike for][and see section~\ref{sec:future}]{Poggio2022}. 

\begin{figure}
	 \includegraphics[width=1\columnwidth]{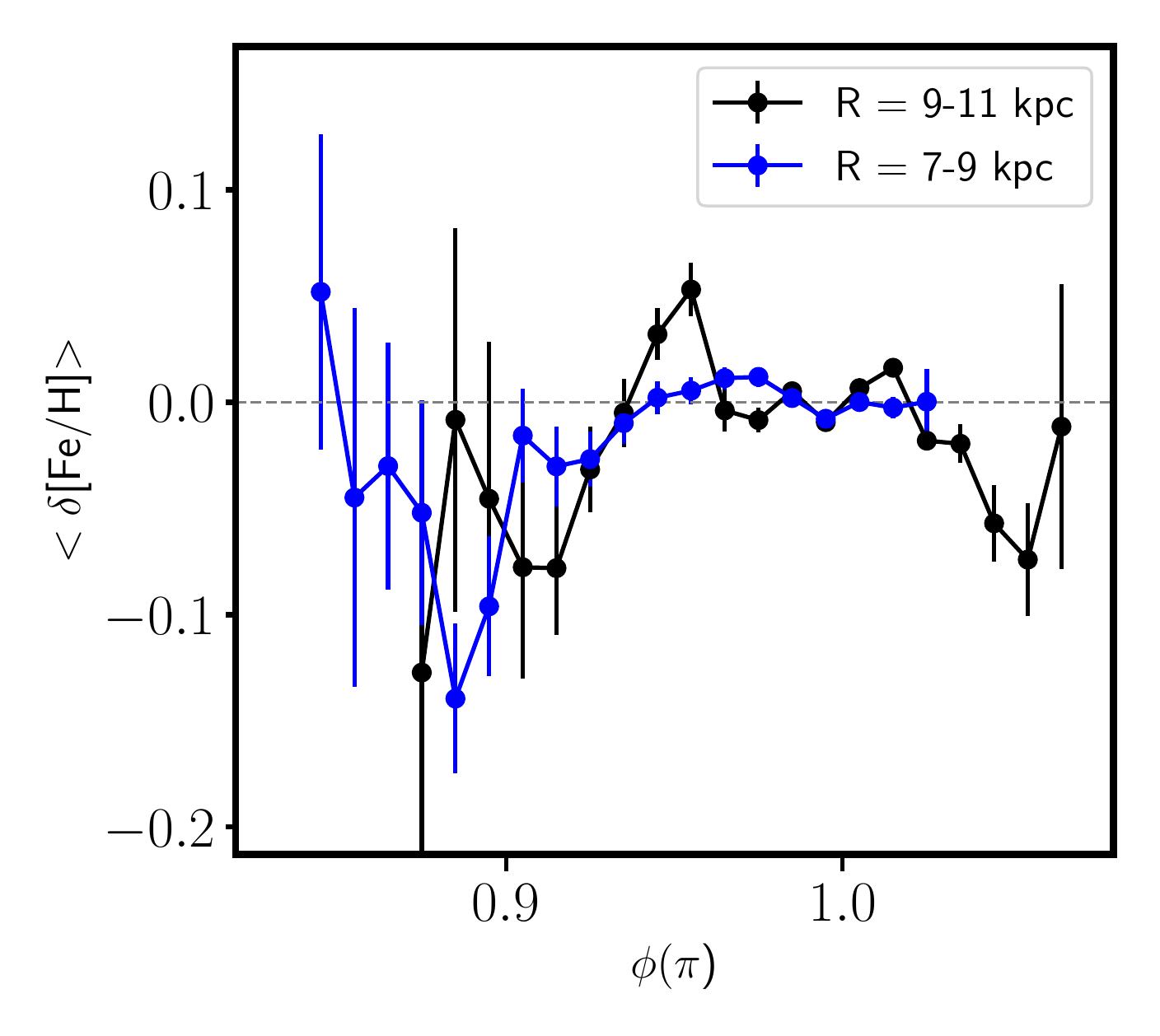}
	\caption{The average metallicity deviation at fixed radius, $<\delta$\feh$>$, as a function of azimuthal angle ($\phi$ in radians) for stars in the outer (9 $<$ R $<$ 11~kpc; black line) and in inner (7 $<$ R $<$ 9~kpc; blue line) Galactic disk. If all stars at a constant radius (regardless of azimuthal angle) had the same metallicity $<\delta$\feh$>$ would be zero. We find average metallicity deviation at fixed radius as high as $\sim$0.10~dex, with slightly larger variations the outer disk (black line).   }   
	\label{fig:dFEH}
\end{figure}

Similar to \cite{Khoperskov2018} and \cite{Carr2022}, we further quantify the azimuthal variations by subtracting off  each star's metallicity by the average metallicity of stars at a comparable shared radius\footnote{We define stars that have Galactocentric radii within 0.20~kpc of each other to have a shared radius. This is motivated by \cite{Carr2022} and given the known radial metallicity gradient, the average metallicity within this annulus should not change by more than 0.01~dex.}. This quantity, which we refer to as the average metallicity deviation at fixed radius, is defined as $<\delta$\feh$>$ = [Fe/H](R, $\phi$) -  $<$ [Fe/H](R)$>_\phi$. In this equation,  [Fe/H](R, $\phi$)  represents the metallicity of stars at a given Galactocentric radius and azimuthal angle ($\phi$) while the $<$ [Fe/H](R)$>_\phi$ term represent the average metallicity of stars at the same Galactocentric radius but across all azimuthal angles at that fixed R. In Figure~\ref{fig:dFEH} we show the average metallicity deviation at fixed radius, $<\delta$\feh$>$, as a function of azimuthal angle ($\phi$ in radians) for stars in the outer (9 $<$ R $<$ 11~kpc) disk (black line) and the inner (7 $<$ R $<$ 9~kpc) disk (blue line).  The error bars represent the standard error of the $<\delta$\feh$>$ at a given azimuthal angle. 

We find that the average metallicity deviation at a fixed R varies as a function of azimuthal angle at the level of up to $\sim$0.10~dex. Interestingly, this size of azimuthal variation (along with its pattern with azimuthal angle), is similar to what is expected ( $<\delta$\feh$> \sim$ 0.07~dex) theoretically if induced by radial mixing of stars due to perturbations from a Sagittarius-like  dwarf galaxy \cite[e.g.,][]{Carr2022}. However, we note that other studies  \citep[e.g.,][see their Figure 6]{Khoperskov2018} have found that azimuthal variations could also arise due to secular processes with and without assuming an initial radial metallicity gradient. In future work, we will compare more directly to simulations but the small azimuthal coverage makes those comparisons currently difficult.  

\subsection{Vertical Metallicity Gradient} \label{subsec:vertical}
\begin{figure}
	 \includegraphics[width=1\columnwidth]{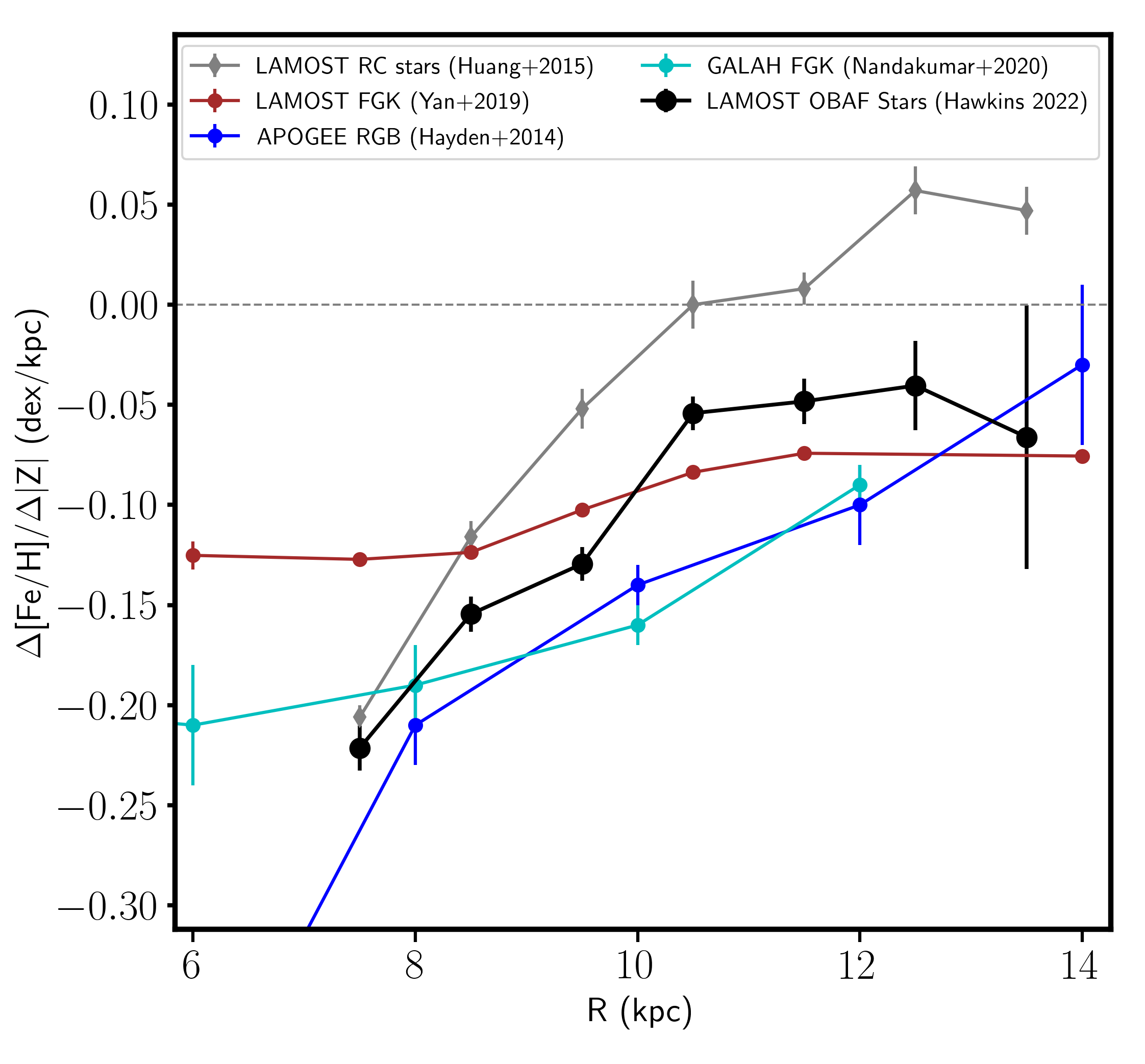}
	\caption{The vertical metallicity gradient, \Vgrad, as a function of Galactocentric radius (R) for the OBAF-type stars (black circles/line) used in this study. For comparison, we also show vertical metallicity gradient measured using: GALAH FGK-type stars \protect\citep[][cyan circles/line]{Nandakumar2020},  LAMOST survey red clump stars \protect\citep[][grey diamonds/line]{Huang2015},  LAMOST survey FGK-type stars  \protect\citep[][brown cirlces/line]{Yan2019},  APOGEE red giant branch stars  \protect\citep[][blue cirlces/line]{Hayden2014}. The Galactic vertical metallicity gradient becomes shallower (i.e., \Vgrad\ becomes less negative and closer to zero) with increasing distance from the center of the Galaxy. }   
	\label{fig:Vgrad}
	
\end{figure}

In addition to deriving the radial metallicity gradient and the azimuthal variations in the metallicity distribution, we also use the LAMOST OBAF-type stars to explore the vertical metallicity gradient. We apply the same procedure, outlined in section~\ref{sec:Methods}, to derive both the radial and vertical metallicity gradients. Previous studies \citep[e.g.][and references therein]{Onal2016} have shown that the vertical metallicity gradient likely varies as a function of Galactocentric radius. Therefore, we split the stars into equally-spaced radial bins with $\sim$1~kpc in size. In order to ensure that we can adequately derive the vertical metallicity gradient we also require a minimum of 500~stars per bin. In Figure~\ref{fig:Vgrad} we show the vertical metallicity gradient derived using the OBAF-type stars observed in LAMOST (black circles). We also tabulate the full linear model fit parameters for each radial bin in Table~\ref{tab:Vgrad}.

We find that the vertical metallicity gradient changes with Galactocentric radius. In the solar neighborhood (7 $<$ R $<$ 9~kpc) the vertical metallicity gradient ranges between --0.22 $<$ \Vgrad\ $<$ --0.15~dex/kpc.  The vertical gradient then becomes significantly shallower (--0.05~dex/kpc) towards the outer Galaxy. As illustrated in Figure~\ref{fig:Vgrad}, these values for the vertical gradient are largely consistent with other tracers, especially in the solar neighborhood. While there are  some differences in our measured  vertical metallicity gradient compared to others, the overall shallowing of the vertical gradient is seen in nearly all tracers. If more metal-poor thin disk stars are on average older than those which are metal-rich, the negative vertical metallicity gradient would imply that older thin disk stars can be found at larger vertical heights on average. This negative vertical metallicity gradient is therefore expected due to kinematic heating of thin disk population or early star formation in a more vertically extended disk.

\setlength{\tabcolsep}{2pt}
\begin{table}
\caption{Vertical Metallicity Gradient Fit Parameters across All Radial Bins}
\begin{tabular}{llllllll}
\hline\hline
R& \Vgrad & $\sigma$\Vgrad & $b_Z$ & $\sigma b_Z$ & $\lambda_Z$ & $\sigma\lambda_Z$ & N\\
(kpc) & (dex/kpc) & (dex/kpc) & (dex)& (dex) & (dex) & (dex)& \\
\hline
7.5 & -0.221 & 0.011 & -0.186 & 0.004 & 0.299 & 0.002 & 8708 \\
8.5 & -0.155 & 0.009 & -0.235 & 0.002 & 0.292 & 0.001 & 37982 \\
9.5 & -0.129 & 0.008 & -0.299 & 0.002 & 0.319 & 0.001 & 42871 \\
10.5 & -0.054 & 0.008 & -0.398 & 0.004 & 0.339 & 0.002 & 23880 \\
11.5 & -0.048 & 0.011 & -0.464 & 0.006 & 0.367 & 0.003 & 11320 \\
12.5 & -0.040 & 0.022 & -0.506 & 0.014 & 0.418 & 0.006 & 3213 \\
13.5 & -0.066 & 0.066 & -0.492 & 0.043 & 0.515 & 0.017 & 525 \\
\hline\hline
\end{tabular}
NOTE: Column 1 tabulates the center of the radial bin (in kpc). The derived vertical metallicity gradient slope and its uncertainty can be found in columns 2 and 3, respectively. The intercept of the linear model and its uncertainty are tabulated in columns 4 and 5, respectively. The intrinsic scatter around the linear \feh-|Z| relation and its uncertainty can be found in columns 6 and 7, respectively. 
\label{tab:Vgrad}
\end{table}

\section{Exploring the Azimuthal Structure in Metallicity Distribution in the Galactic Disk with Gaia DR3}\label{sec:future}
\begin{figure*}
	 \includegraphics[width=2.1\columnwidth]{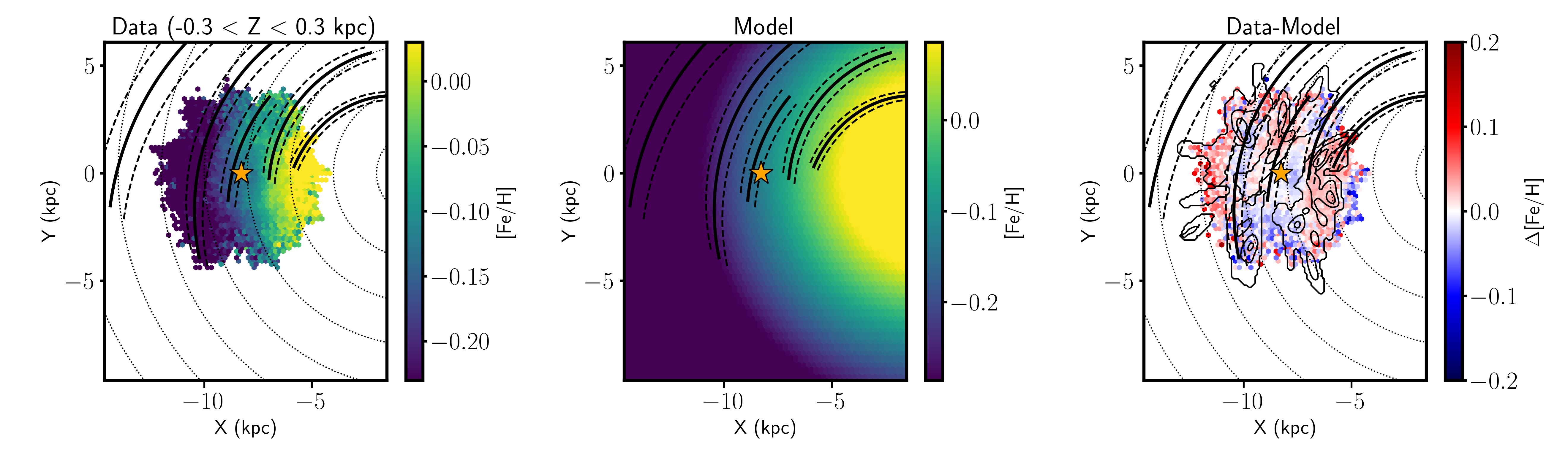}
	\caption{Left: The X-Y projection for the $\sim$480 000 stars in the complementary Gaia DR3 gradient sample of giant stars with 0.00 $<$ |Z| $<$ 0.30~kpc color-coded by the median metallicity derived with using the Gaia RVS spectra. Middle: The X-Y projection of the metallicity gradient defined by the best fitting model.  Right: The X-Y projection of the residual (color coding) between the observed metallicity (right panel) and the predicted metallicity (middle panel). Similar to Figure~\ref{fig:resid}, in all panels the position of the spiral arms (as determined by high mass star forming regions) from \protect\cite{Reid2014} are shown as black lines (both dashed and solid). In the right panel, we also show the position of the spiral arms (black contour), as determined by the upper main sequence, from \protect\citep{Poggio2021}. Interestingly, the residuals (right panel) show azimuthal variations that track with the location of the spiral arms of \protect\cite[shown as black contours][]{Poggio2021}. } 
	\label{fig:resid_Gaia}
\end{figure*}

With the radial and vertical metallicity gradients measured and the azimuthal variations quantified for the LAMOST hot star sample (section~\ref{sec:result}), here we briefly discuss the future prospects of constraining azimuthal structure in the chemical maps of the Galactic disk using Gaia DR3. We are motivated to explore the use of Gaia DR3 data because it is all sky and encompasses a larger swath of azimuthal angles than the LAMOST dataset. In this section, we aim to simply explore whether the azimuthal variations can also be found in the the Gaia DR3 Gradient sample (outlined in section~\ref{subsec:DR3RVS}). Initially we used this dataset in section~\ref{sec:Methods}, to validate our method to derive the radial and vertical metallicity gradients. Using our method, we find that the radial gradient is consistent with those reported in \cite{GaiaCC22}. This effectively validates that our method give reasonable results for the radial and vertical gradients. 

However, with these data, we can also constrain whether there is azimuthal structure using the same procedures as the LAMOST dataset (see section~\ref{subsec:azimuthal}). We start by taking a subsample of  $\sim$480 000 stars in the complementary Gaia gradient sample of giant stars with 0.00 $<$ |Z| $<$ 0.30~kpc and deriving the radial metallicity gradient. In the left panel of Figure~\ref{fig:resid_Gaia}, we show the X-Y projection of the metallicity distribution in the Galactic disk for these $\sim$480 000 giant stars, which are 0.30~kpc from the mid-plane of the Galactic disk. The middle panel of Figure~\ref{fig:resid_Gaia} represents the best fitting radial metallicity gradient (i.e., the expected \feh\ at a given metallicity based on the data). In the right panel of Figure~\ref{fig:resid_Gaia}, we show the residuals between the observed and predicted metallicities (i.e., the observed metallicity subtracted from the predicted metallicity of each star given its Galactocentric radius). For reference, in all panels we show the location of the spiral arms as identified using high-mass star-forming regions \citep[shown as black lines,][]{Reid2014}. We also show in right panel of  Figure~\ref{fig:resid_Gaia}, the location of the spiral arms, as determined by the over density of the upper main sequence stars  \citep[shown as black contours][]{Poggio2021}. {\it Remarkably, we find (1) there is azimuthal structure in the residuals and it is at nearly the same levels ($\sim$ 0.10~dex) as the LAMOST sample, and (2) the residuals show azimuthal structure that seems to be close to the expected location of the spiral arms \citep[e.g.][]{Poggio2021}.} This result indicates that giant stars near the spiral arms (as traced by the upper main sequence) may contain slightly more metals than those which lie in between the arms. This is also consistent with \cite{Poggio2022} who find metal-enhanced elongated structures located in near the spiral arms in the Galactic disc with a slightly different giant subsample of the Gaia DR3 data. We note that the location of the spirals as traced by high mass star formation regions \citep[e.g.][]{Reid2014}, is slightly different than those traced by the upper main sequence \citep[similar to the result from][]{Poggio2021}. This result carries the same limitations as the OBA sample above, namely that the abundances derived in Gaia are computed assuming a range of assumptions (e.g., local thermodynamic equilibrium), and the methodology outlined above (in Section~\ref{sec:Methods}), explicitly models the data under a Bayesian formalisms assuming an intrinsic dispersion around the metallicity gradient independent of location.  Further interpretation of these results will require chemo-dynamical modeling \citep[e.g.][]{Dimatteo2013, Khoperskov2018,Fragkoudi2018, Carr2022}. We stress that this is just a first exploration of the azimuthal variations in metallicity in Gaia data and that followup work will be required to fully understand and map these large-scale chemical structures in the Galactic disk.

\section{Summary} \label{sec:summary}
The industrial revolution in astrometric and spectroscopic surveys of the Milky Way has enabled a large-scale mapping of its disk. In particular these new large-scale datasets have allowed us to chemically map our Galaxy, leading to a refinement of our understanding of the way the Galaxy is structured. In particular, these data have been routinely used to show that our Galaxy has negative radial and vertical metallicity gradients. However, the presence (or lack thereof) and size of azimuthal variations in the metallicity distribution of stars observed in our Galactic disk has been not well explored despite numerous theoretical studies.  Therefore, in this work we aim to use a sample of relatively hot and young stars from the LAMOST survey along with a complementary sample of metallicities from the Gaia DR3 radial velocity spectrometer (see section~\ref{sec:data}) to constrain: (1) the radial gradient in the Galactic disk, (2) the vertical gradient in the Galactic disk, and (3) the level of any azimuthal variations that may lie on top of the metallicity gradient and how it depends on azimuthal angle. The latter of these will determine whether azimuthal variations in the metallicity distribution follow the spiral pattern in the Milky Way.  In order to do this, we set up a Bayesian framework (see section~\ref{sec:Methods}) to derive the linear radial and vertical metallicity gradient, which accounts for the uncertainties in \feh, R, and Z. 

Our main results (see sections~\ref{sec:result}, \ref{sec:future}) can be summarized as follows :
\begin{enumerate}
\item We find a negative radial metallicity gradient, \Mgrad, which can be as steep as $-0.078$~$\pm$~0.001~dex/kpc for  OBAF-type stars in the Galactic disk observed in the LAMOST survey. This gradient is broadly consistent with previous studies of the metallicity gradient with different tracers (from main sequence to red clump stars) and a wide variety of surveys. The gradient we that we find  here is slightly shallower than those found in the APOGEE survey. 

\item The radial metallicity gradient is steepest in the Galactic mid-plane (with \Mgrad\ = $-0.078$~$\pm$~0.001~dex/kpc) but becomes shallower (and nearly flat) as one goes to a larger vertical extent (\Mgrad\ decreases to $-0.001$~$\pm$~0.004~dex/kpc at 1.3~kpc above the mid-plane). This is consistent with other tracer and studies. 

\item We find that there is {\it resolvable azimuthal structure in deviations away from the radial metallicity gradient.} These deviations can be as large as 0.05--0.10~dex. Excitingly, these azimuthal variations show large scale structure and are predicted to originate due to secular processes \citep[e.g.][]{Dimatteo2013, Khoperskov2018,Fragkoudi2018, Wheeler2021} as well as tidal effects from the passage of the Sagittarius dwarf \citep[e.g.][]{Carr2022}. The extent of azimuthal angles probed in this study is somewhat limited, preventing direct comparisons with the simulation. {\it Interestingly, while the azimuthal variations in the LAMOST OBAF star sample do not track the expected location of the spiral arms (see Figure~\ref{fig:resid}), the complementary Gaia DR3 sample of giants show azimuthal variations in the metallicity distribution that grossly follows the spiral arms (see Figure~\ref{fig:resid_Gaia}).} This suggests that while chemical cartography can reveal the spiral structure of the Milky Way \citep[e.g., see also][]{Poggio2022}, it is dependent on tracer population. Regardless, the presence and size of the azimuthal variations is irrespective of which sample is used. Further exploration on the azimuthal structures will be done in followup studies. 

\item We find a negative vertical metallicity gradient, \Vgrad, which can be as steep as $-0.22$~$\pm$~0.01~dex/kpc for  OBAF-type stars in the Galactic disk observed in the LAMOST survey. This gradient is broadly consistent with previous studies of the metallicity gradient with different tracers. 

\item The vertical metallicity gradient in the solar neighborhood (7 $<$ R $<$ 9~kpc) ranges between --0.22 $<$ \Vgrad\ $<$ --0.15~dex/kpc.  The vertical metallicity gradient becomes shallower in the outer galaxy  (12.5$<$ R $<$ 13.5~kpc) ranging between --0.06 $<$ \Vgrad\ $<$ --0.04~dex/kpc. This is consistent with other tracer and studies. This result is broadly consistent with other tracers, and surveys. 
\end{enumerate}

Taken together, our results show that while vertical and radial metallicity gradient are present in the Milky Way, there are further (azimuthal) complexities buried in the data that must be explored (see section~\ref{sec:future}). We recommend further exploration in the azimuthal structure be carried out controlling for tracer population (i.e., spectral type) and stellar age. The results presented here  illustrate the power in chemical cartography in this new industrial era of Galactic and stellar astronomy. 

\section*{Data Availability }
The data (Gaia DR3 and the hot star sample from the LAMOST survey) underlying this article were accessed from both the Gaia Archive (\url{https://gea.esac.esa.int/archive/}), and the Strasbourg astronomical Data Center (CDS) (\url{https://cdsarc.cds.unistra.fr/viz-bin/cat/J/A+A/662/A66}). The derived data generated in this research will be shared on reasonable request to the corresponding author.

\section*{Acknowledgements}
{\small 
KH thanks the anonymous referee for both thorough and helpful comments that improved the manuscript! KH also thanks Carrie Filion, Chris Carr, Paula Jofr\'e, Kathryn Johnston, Adrian Price-Whelan, Dan Foreman-Mackey, Zack Maas, Catherine Manea, Zoe Hackshaw, and Maddie Lucey for fruitful discussions that helped improve this work. KH acknowledges support from the National Science Foundation grant AST-1907417 and AST-2108736 and from the Wootton Center for Astrophysical Plasma Properties funded under the United States Department of Energy collaborative agreement DE-NA0003843. This work was performed in part at Aspen Center for Physics, which is supported by National Science Foundation grant PHY-1607611. This work was performed in part at the Simons Foundation Flatiron Institute's Center for Computational Astrophysics during KH's tenure as an IDEA Fellow. 

This project was developed in part at the Gaia Fête, hosted by the Flatiron Institute Center for Computational Astrophysics in 2022 June. This project was developed in part at the Gaia Hike, a workshop hosted by the University of British Columbia and the Canadian Institute for Theoretical Astrophysics in 2022 June.
 
 The following software and programming languages made this research possible: topcat (version 4.4; \citealt{TOPCAT}); Python (version 3.7) and accompany packages astropy (version 2.0; \citealt{Astropy}), scipy \citep{scipy2020}, matplotlib \citep{matplotlib},  NumPy \citep{numpy}, and emcee \citep{emcee}. This research has made use of the VizieR catalog access tool, CDS, Strasbourg, France. The original description of the VizieR service was published in A\&AS 143, 23.

This work has made use of data from the European Space Agency (ESA)
mission {\it Gaia} (\url{https://www.cosmos.esa.int/gaia}), processed by
the {\it Gaia} Data Processing and Analysis Consortium (DPAC,
\url{https://www.cosmos.esa.int/web/gaia/dpac/consortium}). Funding
for the DPAC has been provided by national institutions, in particular
the institutions participating in the {\it Gaia} Multilateral Agreement.

 This work has made use of data products from the Guoshoujing Telescope
(the LAMOST). LAMOST is a National Major Scientific Project built by the
Chinese Academy of Sciences. Funding for the project has been provided by
the National Development and Reform Commission. LAMOST is operated and
managed by the National Astronomical Observatories, Chinese Academy of
Sciences.

}

\bibliography{bibliography}
\bsp	
\label{lastpage}
\end{document}